\pdfoutput=1
\documentclass[12pt]{iopart}

\makeatletter
\def\@fnsymbol#1{\ifcase#1\or * \or  $+$ \or  \$ \or \#  \or \dag \or \ddag \or
$\mathsection$ \or $ \mathparagraph$ \or $\|$  \or
\textordfeminine \or \textbullet   \or ** \or $++$ \or  \$\$ \or
\#\#  \or \dag\dag \or \ddag\ddag \or $\mathsection\mathsection$
\or $ \mathparagraph\mathparagraph$ \or $\|\|$  \or
\textordfeminine\textordfeminine \or \textbullet \textbullet \or
*** \or $+++$ \or  \$\$\$ \or \#\#  \or \dag\dag \or \ddag\ddag
\or $\mathsection \mathsection\mathsection$ \or $ \mathparagraph
\mathparagraph\mathparagraph$ \or $\|\|\|$  \or
\textordfeminine\textordfeminine\textordfeminine \or
\textbullet\textbullet\textbullet \or \else \@ctrerr\fi}
\makeatother

\usepackage{color}
\usepackage{times}
\usepackage{graphicx}
\usepackage{fancyhdr}
\usepackage{float}

\renewcommand{\today}{\number\day\space\ifcase\month\or
  January\or February\or March\or April\or May\or June\or
  July\or August\or September\or October\or November\or December\fi
  \space\number\year}

\def\be{\begin{equation}}
\def\ee{\end{equation}}
\def\bi{\begin{itemize}}
\def\ei{\end{itemize}}
\def\ben{\begin{enumerate}}
\def\een{\end{enumerate}}

\newcommand\ligodoc{P080029-v3}

\begin{document}

\title{Feasibility of measuring the Shapiro time delay over meter-scale distances
}

\author{S Ballmer$^{1,2,3}$, S M\'{a}rka$^4$ and P Shawhan$^5$}
\address{$^1$Syracuse University, Syracuse, NY 13244, USA}
\address{$^2$TAMA - National Astronomical Observatory of Japan, Tokyo 181-8588, Japan}
\address{$^3$LIGO - California Institute of Technology, Pasadena, CA  91125, USA}
\address{$^4$Columbia University in the City of New York, New York, NY 10027, USA}
\address{$^5$University of Maryland, College Park, MD 20742, USA}

\ead{sballmer@ligo.caltech.edu, smarka@phys.columbia.edu, pshawhan@umd.edu}

\begin{abstract}
The time delay of light as it passes by a massive object, first
calculated by Shapiro in 1964, is a hallmark of the curvature of
space-time.  To date, all measurements of the Shapiro time delay have
been made over solar-system distance scales.
We show that the new generation of kilometer-scale laser interferometers
being constructed as gravitational wave detectors, in particular Advanced
LIGO, will in principle be sensitive enough to measure
variations in the Shapiro time delay
produced by a suitably designed
rotating object placed near the laser beam.  We show that such an
apparatus is feasible (though not easy) to construct, present an
example design, and
calculate the signal that would be detectable by Advanced LIGO.  This
offers the first opportunity to measure space-time curvature effects on
a laboratory distance scale.

\end{abstract}


\submitto{\CQG}

\section{Introduction}

The general theory of relativity asserts that
the familiar force of gravity is fundamentally a manifestation of the
geometry of space-time,
which is given curvature by massive objects.
Many measurements of observable quantities which are influenced by
space-time curvature have confirmed the validity of the Einstein
Equivalence Principle---the foundation on which all metric theories of
gravity are based---and the apparent correctness of general relativity
as the specific metric theory~\cite{TEGP}.  Nevertheless, because of
the importance of space-time and gravity to our understanding of the
universe, opportunities for improved precision and new tests continue
to be sought~\cite{WillLivingReview}.

The most stringent tests of general relativity have been obtained from
measurements of the time delay of an electromagnetic wave as it
passes by a massive
object.  Although this effect is a natural consequence of the equivalence
principle and the curvature of space, it was not called out
by Einstein and was first described by Irwin I. Shapiro in 1964 in the
context of timing radar pulses reflected from Venus and Mars at
superior conjunction~\cite{ShapiroPRL}.  The extra one-way
time delay due to general relativity is
\begin{equation} \label{eq:deltat}
 \delta t = 2 \frac{ G M_\odot}{c^3} \ln\left(\frac{4 r_1 r_2}{d^2}\right)
  \, ,
\end{equation}
where $M_\odot$ is the mass of the Sun, $c$ is the speed of light, $G$ 
is the gravitational constant, $r_1$ and $r_2$ are the
orbital radii of the Earth and the target planet, and $d$ is the
distance of closest approach of the radar beam to the center of the
Sun (assumed here to be much smaller than $r_1$ and $r_2$).
For a radar reflection from Venus at the edge of the Sun, this
amounts to a one-way time delay of $0.12$~ms.

The time delay may be different in alternative theories of gravity.
In the Parametrized Post-Newtonian (PPN) formalism for arbitrary
metric theories~\cite{WillNordtvedtPPN,TEGP}, the time delay is
\begin{equation} \label{eq:deltatGR}
 \delta t = (1+\gamma) \frac{G M}{c^3} \ln\left(\frac{4 r_1 r_2}{d^2}\right)
  \, .
\end{equation}
In the $(1+\gamma)$ factor, the 1 reflects the gravitational red-shift
and is fixed for any metric theory of gravity, while the $\gamma$
measures the net curvature of space produced by the mass along
the path of the light or radio waves~\cite{TEGP}.
For general relativity, $\gamma=1$, reproducing Shapiro's
result.  The possible deviation of $\gamma$ from unity has been
limited by radar ranging to Venus, Mars, and spacecraft, as well as by
timing the radio pulses from isolated and binary
pulsars~\cite{TaylorWeisberg,WillGR75,1998ApJ...505..352S,2006Sci...314...97K};
see~\cite{WillLivingReview} for a more recent summary of measurements.
The most precise measurement was made using the Cassini spacecraft on
its way to Saturn with a Doppler technique,
yielding $\gamma$$-$$1 = (2.1 \pm 2.3) \times
10^{-5}$~\cite{CassiniTest}.
The same $(1+\gamma)$ factor governs the
angular deflection of light passing near the Sun, and
very-long-baseline radio interferometry (VLBI) has been used to
measure $\gamma$$-$$1 = (-1.7 \pm 4.5) \times 10^{-4}$~\cite{GammaVLBI}.
Lunar laser ranging also is sensitive to $\gamma$ in combination with
other PPN parameters~\cite{LLRreview}.

The Sun produces a measurable time delay at its edge despite
the fact that most of its mass is located hundreds of thousands of
kilometers away from the path of the radio beam.  We may wonder
whether smaller objects produce time
delays for smaller separation distances according to the same model,
{\it i.e.}\ the same effective value of $\gamma$.  The PPN formalism does not
allow for such a scale dependence, but there is no experimental
evidence one way or the other for short distance scales.  What time
delay magnitude could we produce on a laboratory scale using a $\simeq$10-ton
mass, say?  As a back-of-the-envelope calculation, utilizing
\eref{eq:deltatGR} with $M$ = $10^4$~kg,
$r_1=r_2=1$~km, and $d=0.50$~m, we estimate that $\delta t = 8.2 \times
10^{-31}$~s.  Remarkably, variations in that time delay due to
{\em changes} in $d$ can be
detected using a suitable laser interferometer---for
instance, one of the Advanced LIGO~\cite{adligo} gravitational-wave detectors---as
we will discuss in section~\ref{sec:LaserInterferometers} below.

\begin{figure}[!t]
\begin{center}
\includegraphics[width=0.8\columnwidth]{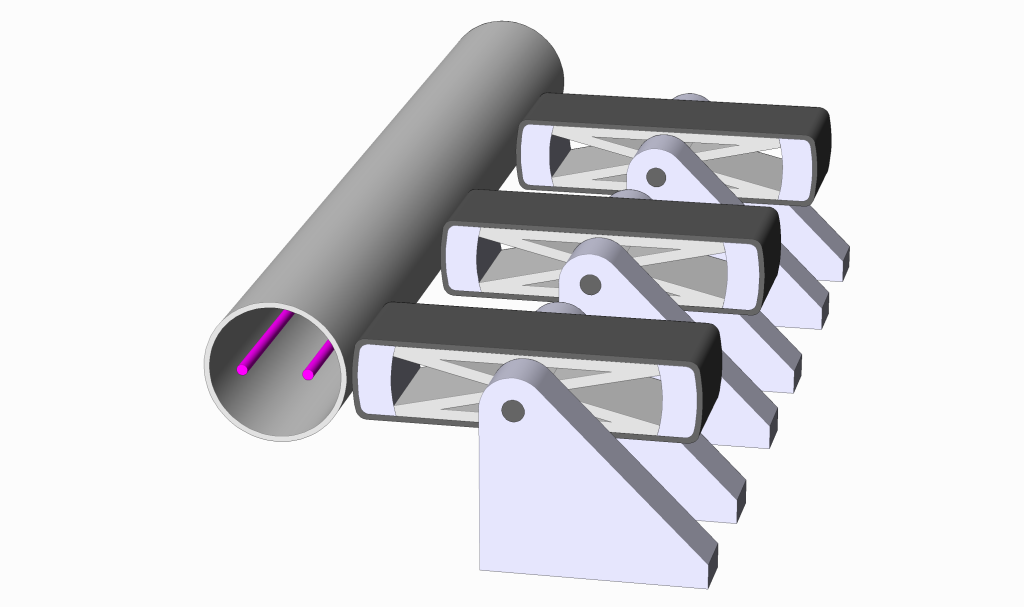}
\caption{To produce a time-varying Shapiro time delay in one arm of a
large interferometer, we envision placing several rotating masses next
to the evacuated tube which encloses the laser beam.  This figure
shows the side-by-side beams of the two interferometers at the LIGO
Hanford Observatory.
The axis of rotation is parallel to the laser beams.
Each section of the rotating mass system is supported by a
bearing assembly in a separate vacuum enclosure (not shown);
they rotate synchronously to produce a
coherent effect on the laser beam.}
\label{fig:ConceptualDrawing}
\end{center}
\end{figure}

If the distance from the mass to the beam changes from $d_{\rm far}$
to $d_{\rm near}$, the resulting change in time delay is
\begin{equation} \label{eq:deltadeltat}
 \Delta \delta t = (1+\gamma) \frac{2 G M}{c^3} \ln\left(\frac{d_{\rm far}}{d_{\rm near}}\right)
  \, .
\end{equation}
Thus, we can spin a massive object near
the interferometer beam to modulate the time delay and then
coherently integrate the signal over a long time period.
Figure~\ref{fig:ConceptualDrawing} is a sketch which illustrates this
basic concept.
In section~\ref{sec:RotatingMassDesign},
we discuss the design considerations for the
rotating mass and present a specific example design which is sufficient to
make an interesting measurement of the time delay, as described in
section~\ref{sec:SignalAnalysis}.
We briefly examine several possible systematic effects and
implementation issues in sections~\ref{sec:Systematics}
and \ref{sec:Engineering} and then conclude with a
discussion of how future interferometers could make better
measurements possible.

\section{Laser Interferometers}
\label{sec:LaserInterferometers}

Direct detection of gravitational waves~\cite{snomass2001} is the
primary goal of modern interferometric gravitational wave detectors
such as LIGO~\cite{S5_LIGO_Instrument},
VIRGO~\cite{virgo,NewVirgoStatus},
GEO600~\cite{NewGEOStatus},
TAMA300~\cite{NewTAMAStatus} and
CLIO~\cite{CLIO}
which have successfully collected a few years of high-sensitivity
data so far.
Second-generation gravitational wave detectors now being designed and
constructed---namely Advanced LIGO~\cite{adligo}, Advanced
Virgo~\cite{advirgo} and the Large-scale Cryogenic Gravitational-wave
Telescope (LCGT)~\cite{lcgt}---will greatly extend the
reach for gravitational wave signals.
Each of these detectors is capable of measuring extremely small changes in the
effective length difference of its two perpendicular ``arms'',
as would be induced by a passing gravitational wave.
However, they can also be
viewed as unique ``laboratory'' instruments capable of measuring
induced displacements at the
$\sim 10^{-23}~{\rm m}$ level, equivalent to time delays of
$\sim 3 \times 10^{-32}~{\rm s}$, for integration times of order one year.

This feature opens up new possibilities for fundamental science.
While gravitational wave searches target signals with various
durations, and either accurately-known or unknown waveforms,
precision experiments in which the light travel time in one of the
interferometer arms is modulated at a given frequency in a well-controlled
manner can enjoy all the benefit of sensitive lock-in experimental
techniques and extended integration times.

Advanced LIGO will consist of three $4\textrm{ km}$
interferometric detectors at the two existing LIGO sites:
the Hanford Observatory in Washington State (home to two
detectors) and the Livingston Observatory in Louisiana.
(A proposed alternative plan, to install one of the three Advanced
LIGO detectors at the Gingin site in Australia, leaving just one at
Hanford, is under consideration at the time of this writing.)
The detectors will be installed beginning in 2011 and are expected to
be fully operational around 2015, ultimately reaching an order of
magnitude lower strain noise than the initial LIGO detectors.

\begin{figure}[!t]
\begin{center}
\includegraphics[width=0.8\columnwidth]{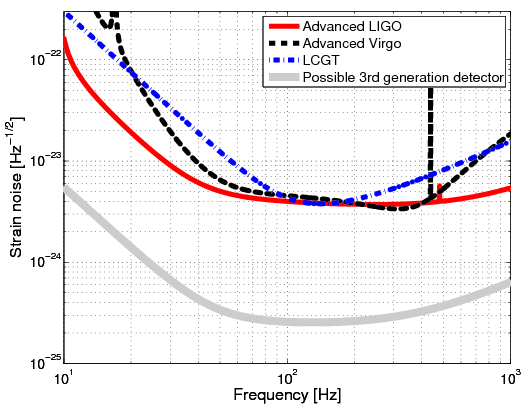}
\caption{Nominal expected noise levels of second-generation
  gravitational wave detectors as a function of frequency, adapted
  from numerical models provided courtesy of the
  LIGO~\cite{ligonoise} and Virgo~\cite{advirgo,virgonoise} projects, and
  published by LCGT~\cite{lcgt}.  The Advanced LIGO curve is for the
  ``zero-detuning, high power'' configuration; see~\cite{ligonoise}
  for a brief summary of alternative configurations and caveats.
  The LCGT curve is for the ``broadband RSE'' configuration.
  The lowest strain noise is obtained for frequencies between about
  70 and 300~Hz.
  Also shown is a representative noise curve for a possible third generation
  detector design~\cite{etnoise}, expected to have an order of magnitude 
  lower noise
  and to have access to lower frequencies.} \label{fig:SensitivityCurves}
\end{center}
\end{figure}

The Advanced LIGO detector design has a beam splitter and suspended
mirrors (test masses) at the ends of its orthogonal arms, as well as
additional partially-transmitting mirrors, to form a power- and
signal-recycled Michelson interferometer with Fabry-Perot arm
cavities.
A gravitational wave induces a time-dependent metric strain
on the detector which changes the relative lengths of the arms.
While acquiring scientific data, feedback to the mirror positions and
to the laser frequency keeps the optical cavities near resonance, so
that interference in the light from the two 4~km long arms recombining
at the beam splitter depends on the difference between the lengths of
the two arms modulated by the signal to be measured. A photodiode
senses the light, and a digitized signal is recorded. Then the data
are calibrated and converted into a strain time series.  The detectors
have a sensitive frequency band extending from a few tens of Hz to a
few kHz, as shown in figure~\ref{fig:SensitivityCurves}, limited at low
frequencies by seismic noise and radiation pressure noise and at high
frequencies by laser shot noise.
Third-generation detectors are envisioned to push the limits of
ground-based interferometry and will likely be limited at low
frequencies by Newtonian gravity-gradient noise~\cite{GravGradient}.

\section{Rotating Mass Design}
\label{sec:RotatingMassDesign}

Since laser interferometers like LIGO have essentially
no sensitivity at DC, it is necessary
to modulate the Shapiro delay by changing
the effective distance between the beam and the relevant mass.
The easiest way to achieve this is a rotating mass system.
In this section we describe one possible design for a device
with two ``arms'' extending from the rotation axis that
would provide a Signal-to-Noise Ratio (SNR) of $\sim$8 for one year
of integration with Advanced LIGO.
(There is no plan at this time to actually construct and
operate such a device, but it could be considered as an addition
to Advanced LIGO or another interferometer in the future.)

As indicated by \eref{eq:deltadeltat},
the geometry of the rotating mass setup enters only
logarithmically into the Shapiro delay, while the mass enters linearly.
The closest effective distance of the mass to beam, $d_{\rm near}$,
is constrained by their finite cross-sectional sizes.
As a consequence it is hard to reduce the required mass by optimizing
the rotating mass geometry.
For the example design we assume a mass of $1.5 \times 10^4~{\rm kg}$
for each of the two arms.

One of the limiting factors is the tensile strength of the material
holding the rotating mass together. This of course favors low spin frequencies,
and therefore requires a laser interferometer with good low-frequency
sensitivity. In the case of Advanced LIGO we would like to have the primary harmonic
at about $50~{\rm Hz}$, well separated from known instrumental lines in the
detector noise spectrum.
Thus, a baseline design with two arms must rotate at $25~{\rm Hz}$.
Note that this is still much slower than the inverse light travel time
for all relevant distances; thus a static calculation of the Shapiro
delay is still appropriate.
A rotating mass system design with more arms would have a lower spin
frequency, but
in order to preserve the size of the variation of
the Shapiro delay, one would have to increase the radius by the same factor.
Since the whole rotating mass system needs to spin in vacuum and it must fit next to the beam
tube we choose $1.5~{\rm m}$ as the maximal practical rotating mass radius and
stay with the conceptually simpler two-arm design and a $25~{\rm Hz}$
spin frequency, even though one
might win a factor of two in required tensile strength by going to 4 arms.

\begin{figure}[!t]
\begin{center}
\includegraphics[width=0.8\columnwidth]{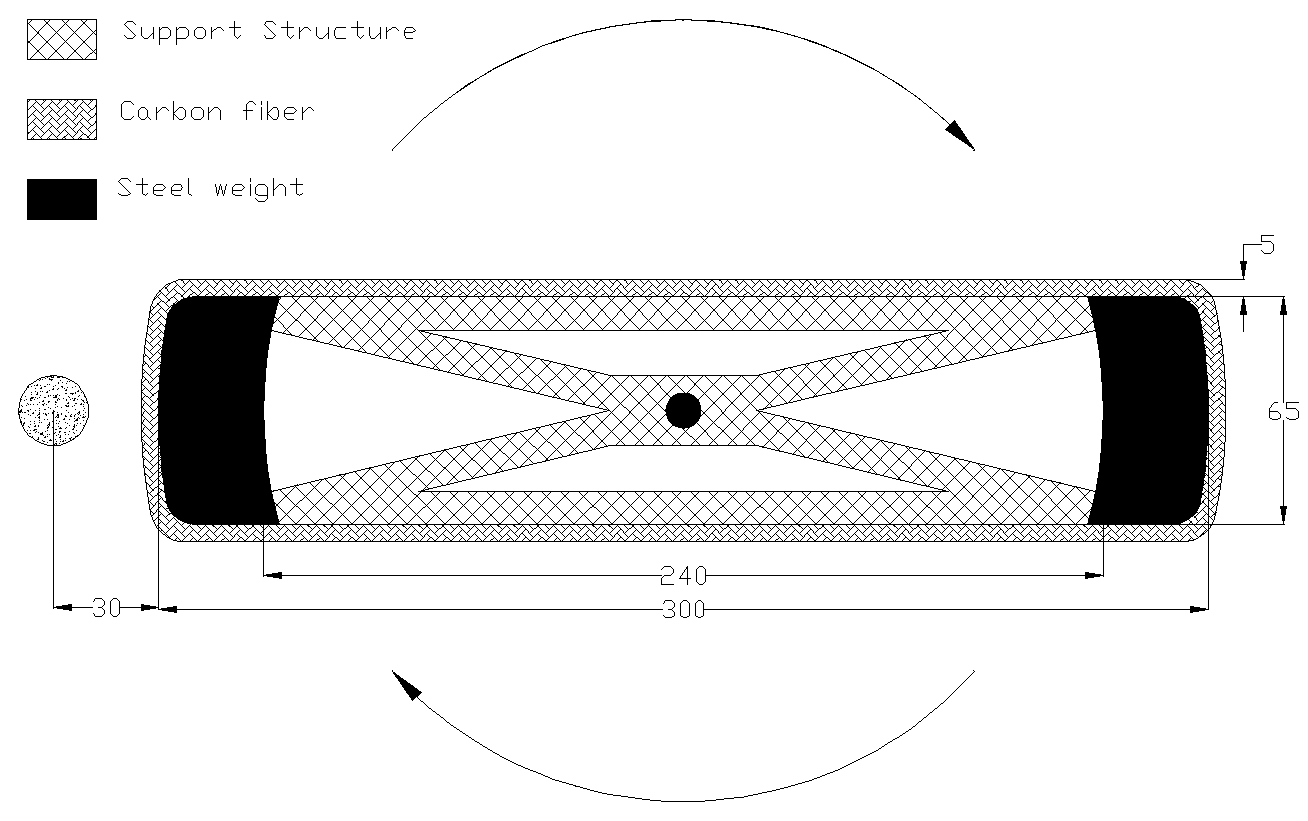}
\caption{Cross-section of the example rotating mass system. All units
  are in cm. The support structure centered on the rotation axis is
  adequate to support the static weight of the steel weights, which
  represent most of the mass of the system. The carbon fiber epoxy
  composite outer envelope provides tensile strength when the system
  is spinning. The center of the laser beam is located 30 cm from the outer edge of
  the masses.} \label{fig:RotatingMass-Model}
\end{center}
\end{figure}

With these parameters the rotating mass structure needs to support a centrifugal force
of about $6 \times 10^{8}$~N. A simple rectangular slab of steel,
tungsten or titanium is not strong enough, and only a fraction of the
mass would pass close to the laser beam.
A better way to build a rotating mass system with the necessary
strength is to concentrate the mass near the ends of the arms, with
a light-weight central support structure for suspension, and an outer shell
of carbon fiber composite. Figure \ref{fig:RotatingMass-Model} shows a sketch
of this rotating mass design, including the key dimensions. With a carbon fiber composite
cross-sectional area of $1~{\rm m}^2$ ($5~{\rm cm}$ thickness on both sides and
$10~{\rm m}$ cumulative rotating mass length along the axis of rotation)
this requires a tensile strength of $600~{\rm MPa}$.
A composite such as HexTow IM9 from HEXCEL corporation~\cite{HEXCEL},
for example, provides a factor of 5 safety margin on this.

Another constraint on the geometry of the setup, specifically the length of the
interferometer arms, comes from the direct gravitational coupling from the rotating mass to
the test masses. A good place to put the rotating mass is near the middle of one
interferometer arm.
The test mass acceleration at the fundamental spin frequency $f_{\rm spin}$ cancels due to the symmetry
of the rotating mass, while the acceleration variation at $2 f_{\rm spin}$ has amplitude
\begin{equation}
a = \frac{15}{2} \frac{G M r^2 (r+d)^2}{ (L/2)^6 }
\label{eq:directAccel}
\end{equation}
where $M$ is the mass of one arm, $r$ is the rotating mass radius, $d$ its minimal separation from the
laser beam, and $L$ is the interferometer arm length.
For a $25~{\rm Hz}$ spin frequency and including the acceleration of both test masses of the interferometer arm
the length change at $f=50~{\rm Hz}$ has amplitude
\begin{eqnarray}
dx &= \frac{240}{\pi^2} \frac{G M r^2 (r+d)^2}{ f^2 L^6 }   \\
   & \approx \left( \frac{384~{\rm meter}}{L} \right)^6  \frac{2 G M}{c^2} \, .
\label{eq:directDispl}
\end{eqnarray}
Since the size of the Shapiro delay is on the order of the Schwarzschild radius
$2 G M / c^2$ of one of the arm masses, divided by $c$,
any laboratory-scale interferometer ($L \simeq 40~{\rm m}$ or less)
would require an extremely precise cancellation mechanism.
For the $L=4~{\rm km}$ arm length of LIGO this coupling is down
by more than 6 orders of magnitude.

\section{Signal Analysis}
\label{sec:SignalAnalysis}

\begin{figure}[!t]
\begin{center}
\includegraphics[width=0.9\columnwidth]{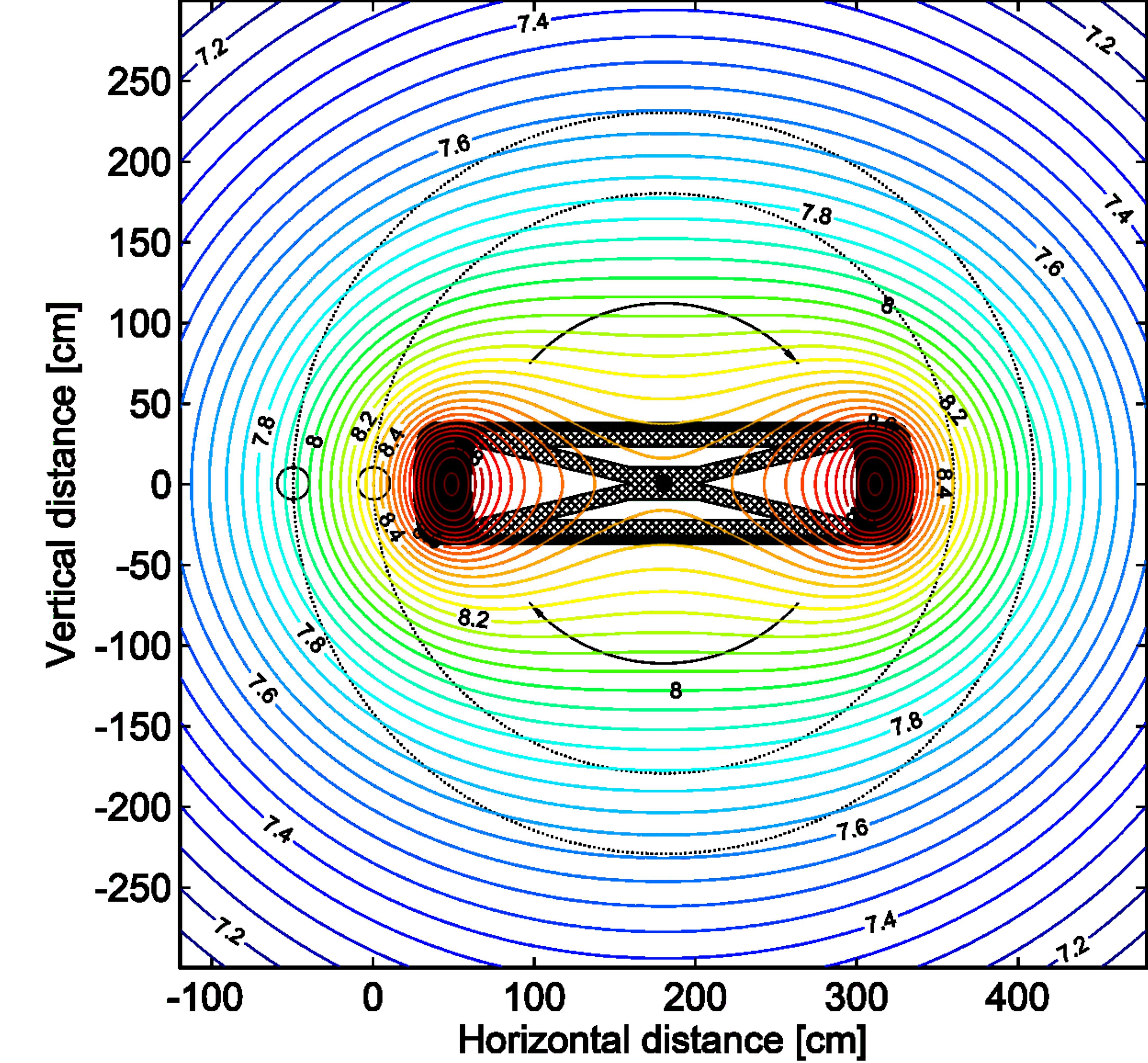}
\caption{Shapiro time delay of the laser beam as a function of the position around the rotating mass. Note that the contour pattern rotates with the mass. The contour labels are in units of 100 yoctometers ($10^{-22}$~m). The small black circles illustrate the relative size and position of the laser beams at the time of the closest approach. The large dotted circles trace out the positions of the centers of the laser beams relative to the rotating mass for a full revolution for the specific configuration discussed. The paths of the dashed circles through the isochrones map out the curves shown in figure~\ref{fig:timeseries}.}
\label{fig:isochrones}
\end{center}
\end{figure}

For a given rotating mass, such as the one described in the previous section,
it is straightforward to numerically calculate the gravitational
potential $U$ at every point in its vicinity.  A light ray which
passes near the rotating mass experiences a total time delay which is simply
given by
\begin{equation}
  \delta t = \int (1+\gamma) \, U \, ds
\end{equation}
where the integral is
along the path of the light ray~\cite{TEGP}.
The light also experiences angular deflection during its passage,
but for this system the deflection
is only of order $\sim$$10^{-22}$~rad.  The slightly increased length of
the path due to this deflection is quadratic in the angle, and thus has a
negligible effect on the time delay calculation.
Therefore we may take the light
path to be straight in the $\hat{z}$ direction (along the laser beam)
and calculate the
integrated time delay as a function of its transverse position when it
passes by the rotating mass.  Also, because the potential is additive, the
spatial extent of the rotating mass in the $\hat{z}$ direction does not
matter; we can project all of its mass density to the $x$-$y$ plane.
Figure~\ref{fig:isochrones} shows contours of constant time delay
(isochrones) around our example rotating mass design.

\begin{figure}[!t]
\begin{center}
\includegraphics[width=0.8\columnwidth]{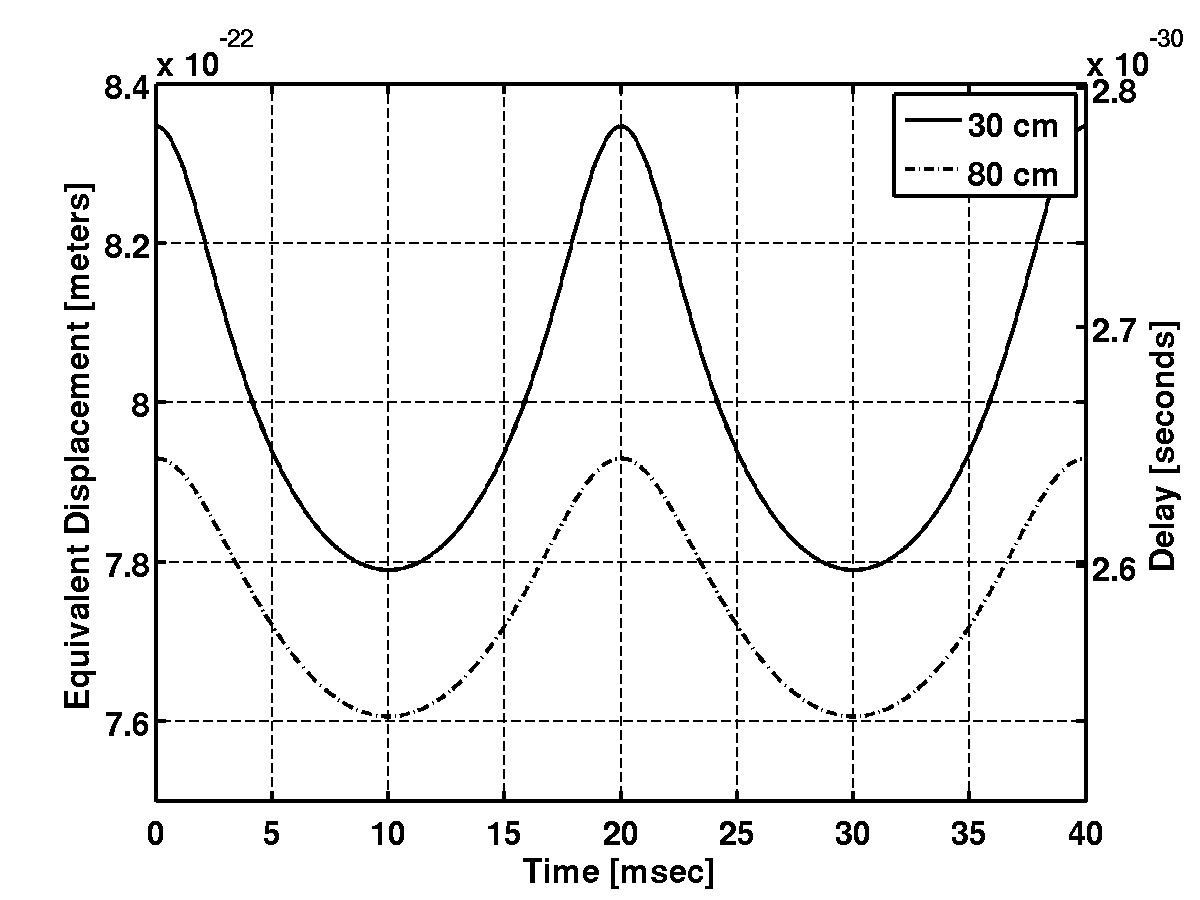}
\caption{Shapiro time delay of the laser beam as a function of the time
for one full rotation of the rotating mass system.
The two traces correspond to laser beam positions $30~{\rm cm}$ and
$80~{\rm cm}$ away from the outer edge of the rotating mass.}
\label{fig:timeseries}
\end{center}
\end{figure}

The time delay experienced by the light in the interferometer beam at
any instant depends on the orientation of the rotating mass.  As seen from the
point of view of the rotating mass, the interferometer beam travels in a
circle around the axis, as illustrated by the dashed lines in
figure~\ref{fig:isochrones} for beams with closest-approach distances of
$30~{\rm cm}$ and $80~{\rm cm}$, representing the two interferometer
beams at LIGO Hanford.  The time delay at the center of the beam
as a function of time is shown in figure~\ref{fig:timeseries} for one
full rotation of the rotating mass.  This is the signal which would appear in
the interferometric data stream.  Because the form of the signal is
known accurately, it may be extracted from the data using lock-in techniques.

\begin{figure}[!t]
\begin{center}
\includegraphics[width=0.8\columnwidth]{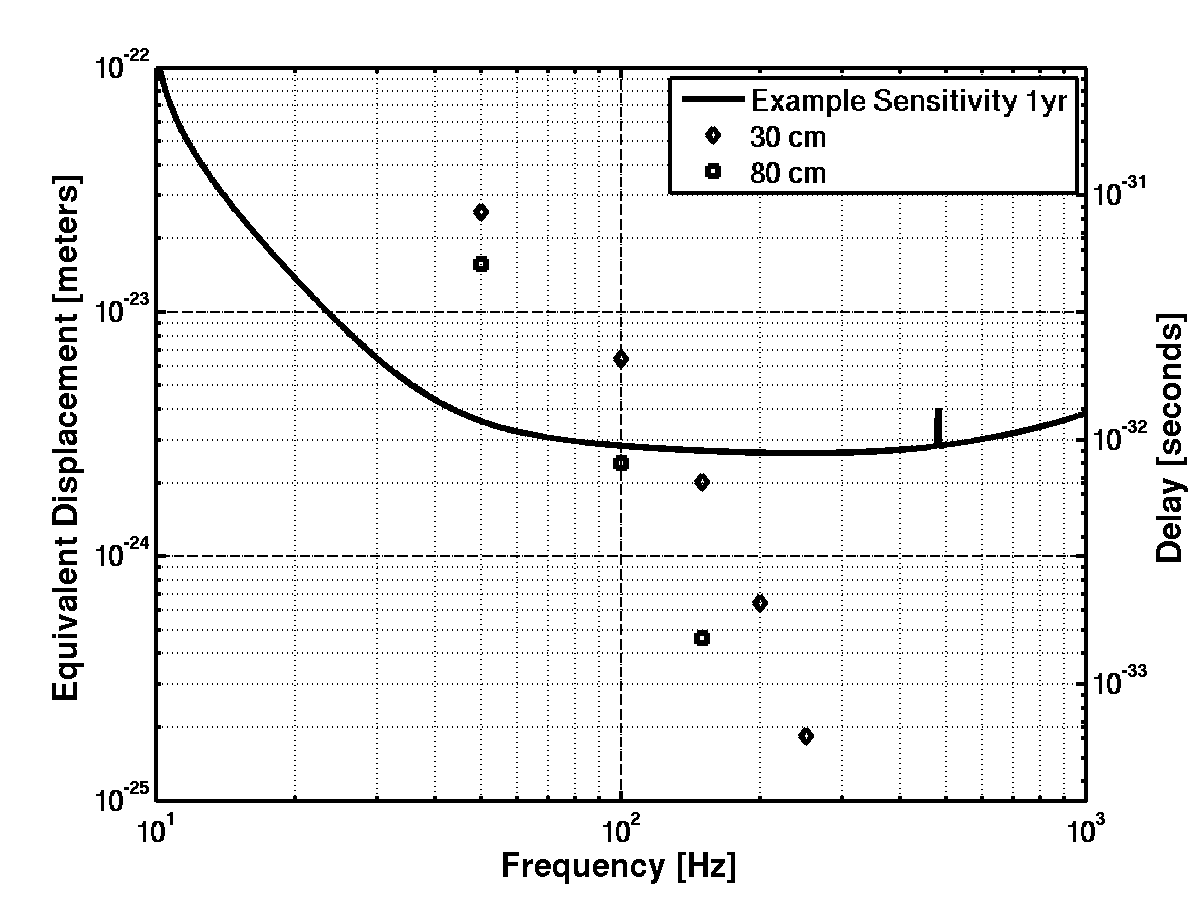}
\caption{Fourier components of the Shapiro time delay signal.
The diamonds and squares correspond to laser beam positions of
$30~{\rm cm}$ and
$80~{\rm cm}$ away from the outer edge of the rotating mass.
Also shown is an example Advanced LIGO sensitivity curve, scaled for 1 year of
integration time.}
\label{fig:freqdomain}
\end{center}
\end{figure}

The signal is not purely sinusoidal because the distances from the
different parts of the rotating mass are comparable to the size of the rotating mass.
Figure~\ref{fig:freqdomain} shows the Fourier components of the signal
(for the nominal spin frequency of 25~Hz)
superimposed on the expected sensitivity curve for
Advanced LIGO assuming an integration time of one year.  Based on this
comparison, the fundamental frequency component at 50~Hz should be
detectable with amplitude SNRs of $7.2$ and $4.4$, respectively,
in the closer and farther beams.
The first harmonic (at 100~Hz) should also be detectable in the
closer beam with SNR of $2.3$.
Comparing these different measurements would test the consistency of
the results with the expected form of the signal, while combining them
in quadrature yields a total SNR of $8.7$.

\section{Possible systematic uncertainties}
\label{sec:Systematics}

To interpret the measured signal as being due to the Shapiro time
delay, we must have sufficiently accurate knowledge of the geometry of
the apparatus and must be able to rule out other possible systematic
effects which could affect and/or fake the Shapiro delay signal.  However, given that we only
expect an SNR of $\sim$8 above the instrument noise, we only need
to limit these systematic effects to a few percent of the expected
signal amplitude.

First, we must know the separation between the rotating masses and the
interferometer beam to within a few percent, {\it i.e.}\ perhaps 1~cm.
The rotating mass system parts can be machined and assembled with much tighter
tolerances, and also measured precisely after assembly.  When spun up,
the rotating mass system would stretch by about $5~{\rm mm}$ due to the extreme
tension on the carbon fiber composite outer layer; this should be taken into account in
the data analysis and may not be known precisely (due to uncertainty
in the bulk modulus of the carbon fiber composite) but would be a small
correction.  Knowledge of the interferometer beam position within its
vacuum tube would be a greater challenge, but can probably be
determined to 1~cm or better. The Advanced LIGO alignment system should
keep the position of the beam stable to much better than 1~mm over the
course of a year-long data run.  There would also be a horizontal
gradient in the time delay over the area of the interferometer beam,
but the intensity-weighted mean delay (over the Gaussian beam profile)
would be well-determined and the gradient is not expected to cause any
other problems.
Additionally, the diameter of the laser beam's vacuum tube may have to
be reduced locally to allow the required close approach of the
rotating mass to the laser beam.

It may be advantageous to minimize the length of beam tube devoted to
the rotating mass system by placing synchronized rotating units on
both sides of the beam tube.  This also would reduce the time delay
gradient over the area occupied by the laser beam.
In the case of LIGO Hanford, rotating mass systems on both sides would
produce equal time delays in both interferometers (but would only
partly reduce the time delay gradient across each laser beam).

We must also consider parasitic coupling of the rotating mass system's motion to the
interferometric signal through mechanisms other than the time delay.
As mentioned in section~\ref{sec:RotatingMassDesign}, direct gravitational coupling to the test
masses would be negligible due to the length of the interferometer
arms.  Electromagnetic interference from the motor driving the rotating masses
should be negligible due to being so far away from the sensing
electronics. Additionally one can use a gear so that the motor runs
at a different frequency.

Another concern is local
vibration of the interferometer beam's vacuum tube, directly excited
by the time-varying gravitational field of the rotating mass system.
This vibration would have a
significant component at twice the spin frequency and could couple
into the interferometric signal by modulating light scattered off the
walls of the beam tube.

Baffles placed at an
appropriate distance can sufficiently reduce the effects of light
scattering.  These baffles must be mechanically isolated from vibrations
coming from either the beam tube or the ground.

The residual coupling from such baffles can also be estimated using 
\eref{eq:directDispl}, with an additional scattering factor.
\begin{equation}
dx_{\rm scat} \approx \left( \frac{E_{\rm scat}}{E_{\rm beam}} \right)
\left( \frac{384~{\rm meter}}{L} \right)^6  \frac{2 G M}{c^2}
\label{eq:directDisplScat}
\end{equation}
Thus placing the baffles a few hundred meters away from the rotating mass
should provide sufficient isolation. Secondary scattering should also be
analyzed, but that goes beyond the scope of this paper.

Such a massive rotating mass system, if even slightly unbalanced, would vibrate the
ground as it rotates.  We should be able to tune the balance below the sensitivity level of seismometers at initial
assembly. However, residual vibration might evolve due to aging or slow differential
stretching of the rotating mass's carbon fiber composite layer when it is spun up to very high speeds.
If the vibration is not sufficiently attenuated
as it travels through 2~km of ground, it could shake the sensing
optics and introduce a spurious signal.  However, the vibrational
signal would be generated primarily at the spin frequency and should
be detectable with seismometers near the rotating mass systems and
near the interferometer test masses.  Thus a
vibrational signal carried through the ground would largely be
distinguishable from the time delay signal, which appears at twice the
spin frequency and harmonics thereof.

We also note that any residual false signal due to vibrational coupling is
likely to appear in the data with a phase different from what is
expected from the relativistic time delay signal; this provides a
useful consistency test.

\section{Engineering considerations}
\label{sec:Engineering}

Actual construction and operation of a rotating mass system like the one described in
this paper would present several engineering challenges, but we
believe that they could all be met with existing technologies.  While
detailed solutions are beyond the scope of this exploratory study, we
can at least comment on the major issues.

In the example rotating mass system design, a total mass of
$\simeq$30 tonnes with
a total length of 10~m must be supported reliably and spun with little
friction.  Rather than attempting to do that with a single
assembly 10~m long, we imagine that it would be divided into several
independent sections, each supported by multiple bearings to be able
to tolerate the failure of a bearing.
The sections would be rotated by independent motors.
The rotation of each section would be monitored with a rotary
encoder and fed back to the drive motor to maintain the desired
spin frequency and phase.  The rotation and drive force would also be
monitored for signs of possible drive or
bearing failure.

The kinetic energy of each rotating section would be of order
100--200~MJ, comparable to the energy in commercially available
flywheel units used for energy storage~\cite{CommercialFlywheel}.
Obviously, a structural failure must be avoided at all
costs; the safety of the facility and personnel would be crucial
considerations if this experiment is to be carried out.

\section{Discussion}
\label{sec:Discussion}

In this paper, we have explored the possibility of 
measuring the effects of space-time curvature over laboratory
distance scales for the
first time.  We have found that it is technically feasible to make
this measurement using the Advanced LIGO interferometers.  However, it
is far from easy.  We had to resort to the use of
high-tensile-strength composite materials to obtain a rotating mass system with
sufficient mass and rotation speed to produce a measurable effect with
an SNR of $\sim$8 integrated over a year of data.  This is sufficient
to measure $\gamma$ with a precision of $\sim$25\%, which could
distinguish between the cases $\gamma=1$ and $\gamma=0$, for instance.

Future interferometers could improve this type of measurement,
especially if their low-frequency sensitivity is better~\cite{LFdets}
so that the mass can be rotated at a lower frequency.
In particular, a design study is currently underway for
a future underground ``Einstein Telescope''~\cite{ET} with projected
sensitivity shown by the lowest curve in figure~\ref{fig:SensitivityCurves}.
That detector could
support a Shapiro time delay measurement at a spin frequency of
$\sim$15~Hz that would be an order of magnitude better than the
Advanced LIGO example presented in this paper.
Besides the planned gravitational-wave interferometers,
a dedicated interferometer could perhaps be used to make this
time delay measurement.
Although we cannot hope to approach the precision of the radar ranging
measurements with this technique, it is interesting to be able to test
gravity in an entirely different distance regime.

\section*{Acknowledgments}

LIGO was constructed by the California Institute of Technology and
Massachusetts Institute of Technology with funding from the National
Science Foundation and operates under cooperative agreements PHY-0107417
and PHY-0757058.
The authors are also grateful for the support of the
National Science Foundation under grants PHY-0457528,
PHY-0653421, PHY-0757957 and PHY-0757982;
Columbia University in the City of New York; the University of Maryland;
and the California Institute of Technology.
We are indebted to
many of our colleagues for fruitful discussions,
in particular Daniel Sigg, Rainer Weiss, Rubab Khan and Zsuzsa M\'arka, and to
Yoichi Aso for providing figure~1.
We also thank an anonymous referee for helpful suggestions.
This document has LIGO Document Number LIGO-\ligodoc.


\section*{References}

\begin{thebibliography}{99}
\bibitem{TEGP} Will C M 1993
   {\it Theory and Experiment in Gravitational Physics}
   2nd ed.\ (Cambridge: Cambridge University Press)
\bibitem{WillLivingReview} Will C M 2006 
   {\it Liv.\ Rev.\ Rel.}\ 2006-3
\bibitem{ShapiroPRL} Shapiro I I 1964
   {\it Phys.\ Rev.\ Lett.}\ {\bf 13} 789--91
\bibitem{WillNordtvedtPPN} Will C M and Nordtvedt K Jr 1972
   {\it Astrophys.\ J.}\ {\bf 177} 757--74
\bibitem{TaylorWeisberg} Taylor J H and Weisberg J M 1989
   {\it Astrophys.\ J.}\ {\bf 345} 434--50
\bibitem{WillGR75} Will C M 1990
   {\it Science} {\bf 250} 770--6
\bibitem{1998ApJ...505..352S} Stairs I H, Arzoumanian Z, Camilo F, Lyne A G,
   Nice D J, Taylor J H, Thorsett S E and Wolszczan A 1998
   {\it Astrophys.\ J.}\ {\bf 505} 352--7
\bibitem{2006Sci...314...97K} Kramer M {\it et al} 2006
   {\it Science} {\bf 314} 97--102
\bibitem{CassiniTest} Bertotti B, Iess L and Tortora P 2003
   {\it Nature} {\bf 425} 374--6
\bibitem{GammaVLBI} Shapiro S S, Davis J L, Lebach D E and Gregory J S 2004
   {\it Phys.\ Rev.\ Lett.}\ {\bf 92} 121101
\bibitem{LLRreview} Williams J G, Turyshev S G and Boggs D H 2004
   {\it Phys.\ Rev.\ Lett.}\ {\bf 93} 261101 
\bibitem{adligo} Harry G M (for the LIGO Scientific Collaboration) 2010
   {\it Class.\ Quantum Grav.}\ {\bf 27} 084006
\bibitem{snomass2001} Hughes S A, M\'arka S, Bender P L and Hogan C J 2001
   {\it Proc.\ APS/DPF/DPB Summer Study on the Future of Particle Physics
   (Snowmass 2001)} ed N Graf, eConf C010630, P402, arXiv:astro-ph/0110349
\bibitem{S5_LIGO_Instrument} Abbott B P {\it et al} 2009
   {\it Rep.\ Prog.\ Phys.}\ {\bf 72} 076901
\bibitem{virgo} Acernese F {\it et al} 2008
   {\it Class.\ Quantum Grav.}\ {\bf 25} 184001
\bibitem{NewVirgoStatus} Accadia T and Swinkels B L (for the VIRGO Collaboration) 2010
   {\it Class.\ Quantum Grav.}\ {\bf 27} 084002
\bibitem{NewGEOStatus} Grote H (for the LIGO Scientific Collaboration) 2010
   {\it Class.\ Quantum Grav.}\ {\bf 27} 084003
\bibitem{NewTAMAStatus} Takahashi R {\it et al} (TAMA Collaboration) 2008
   {\it Class.\ Quantum Grav.}\ {\bf 25} 114036
\bibitem{CLIO} Yamamoto K {\it et al} 2008
   {\it J.\ Phys.: Conf.\ Ser.}\ {\bf 122} 012002
\bibitem{advirgo} The Virgo Collaboration 2009
   {\it Advanced Virgo Baseline Design};
   {\it URL}~https://pub3.ego-gw.it/itf/tds/file.php?callFile=VIR-0027A-09.pdf
\bibitem{lcgt} Kuroda K (on behalf of the LCGT Collaboration) 2010
   {\it Class.\ Quantum Grav.}\ {\bf 27} 084004
\bibitem{ligonoise} Advanced LIGO Interferometer Sensing and Control Group 2010;
   {\it URL}~https://dcc.ligo.org/cgi-bin/DocDB/ShowDocument?docid=2974
\bibitem{virgonoise} {\it URL} http://wwwcascina.virgo.infn.it/advirgo/
\bibitem{etnoise} Hild S, Chelkowski S and Freise A 2008
   {\it Preprint} arXiv:0810.0604v2
\bibitem{GravGradient} Hughes S A and Thorne K S 1998
   {\it Phys.\ Rev.\ D} {\bf 58} 122002
\bibitem{HEXCEL} {\it URL} http://www.hexcel.com
\bibitem{CommercialFlywheel} Lazarewicz M and Rojas A 2004
   {\it IEEE Power Engineering Society General Meeting 2004} vol~2 pp~2038--42
\bibitem{LFdets} DeSalvo R 2004
   {\it Class.\ Quantum Grav.}\ {\bf 21} S1145--54
\bibitem{ET} Punturo M {\it et al} 2010
   {\it Class.\ Quantum Grav.}\ {\bf 27} 084007
\end{thebibliography}
\end{document}